# High-throughput Discovery of Topologically Non-trivial Materials using Spin-orbit Spillage

Kamal Choudhary, Kevin F. Garrity, Francesca Tavazza

Materials Science and Engineering Division, National Institute of Standards and Technology, Gaithersburg, Maryland 20899, USA

Corresponding author: Kamal Choudhary (kamal.choudhary@nist.gov)

## Abstract

We present a novel methodology to identify topologically non-trivial materials based on band inversion induced by spin-orbit coupling (SOC) effect. Specifically, we compare the density functional theory (DFT) based wavefunctions with and without spin-orbit coupling and compute the 'spin-orbit-spillage' as a measure of band-inversion. Due to its ease of calculation, without any need for symmetry analysis or dense k-point interpolation, the spillage is an excellent tool for identifying topologically non-trivial materials. Out of 30000 materials available in the JARVIS-DFT database, we applied this methodology to more than 4835 non-magnetic materials consisting of heavy atoms and low bandgaps. We found 1868 candidate materials with high-spillage (using 0.5 as a threshold). We validated our methodology by carrying out conventional Wannier-interpolation calculations for 289 candidate materials. We demonstrate that in addition to $Z_2$ topological insulators, this screening method successfully identified many semimetals and topological crystalline insulators. Importantly, our approach is applicable to the investigation of disordered or distorted as well as magnetic materials, because it is not based on symmetry considerations. We discuss some individual example materials, as well as trends throughout our dataset, which is available at the websites: https://www.ctcms.nist.gov/~knc6/JVASP.html and https://jarvis.nist.gov/.

## Introduction

Topological materials are a new class of quantum materials, which are characterized by surface states induced by the topology of the bulk band structure[1]. In the past decade, there has been a huge interest in the field of topological materials, which has rapidly expanded to include Chern insulators[2], topological insulators (TI)[3,4], crystalline topological insulators (CTI)[5], Dirac semimetals (DS)[6], Weyl semimetals (WS)[7], quadratic band-crossing materials[8] with many other related subclasses and variations. While some topological materials classes exist without spin-orbit coupling, many topological behaviors are related to spin-orbit coupling (SOC) induced band inversion[9]. Despite their huge potential for technological applications such as robust quantum computation[10], only a few examples of most classes of TMs are known, and a workflow/metrology for discovering, characterizing and finally developing a repository of such materials is still in the developing phase. In this work, we use the spin-orbit spillage criteria[11] (discussed later) and Wannier interpolation[12-14] to quickly screen topological materials. The materials of interest in the present work include topological insulators[3,15], Dirac semimetals[16], Weyl-semimetals[17], quadratic band-crossing materials[18], crystalline topological insulators[5,19] and nodal materials[20,21].

Historically, individual TMs have been initially identified[22-24] mainly by intuition, trial and error, heuristic screening[25], and analogy to other materials[26], often with first principles calculations leading the way. Very recently, there have been several efforts to perform a systematic analysis of band structure symmetries to help identify TMs in systems with enough crystalline symmetries such as by Slager et al., Bradlyn et al., Tang et al. and Zhang et al.[27-35]. In this work, we take a complementary approach to the symmetry-based analysis by directly searching for materials with large changes in the occupied wave functions due to including SOC in a first principles calculation. This approach can identify topological band features and phase transitions in materials with low symmetry, or 'accidental' band crossings at k-points with low symmetry. We can apply our search technique to materials with disordered or distorted structures, which are difficult to analyze with symmetry-based techniques, and our search method can easily be extended to systems with broken time-reversal symmetry, like Chern insulators or magnetic Weyl semimetals. As mentioned above, most classes of TMs require SOC, which leads to band-inversion or splitting of otherwise degenerate bands[11,36]. We calculate the **k**-dependent spin-orbit spillage, and then we screen for materials with large spillage at least at one k-point, as an indicator of band inversion. To classify the previously identified materials, we then use Wannier-interpolation to efficiently calculate topological indices. Using this workflow, we discover hundreds of potential TMs. We added SOC and non-SOC band structures as well as their spillage results to the JARVIS-DFT repository. The JARVIS-DFT website contains about 30000 bulk and 800 two-dimensional materials with their density functional theory (DFT)-computed structural, energetics[37], elastic[38], optoelectronic[39] and solar-cell efficiency[40] properties. Finally, we show examples of several specific TMs identified in this search, and we provide a full list of TMs we have identified in the Supplementary Materials.

Changes in bands structures due to SOC are generally ≤ 1eV and are larger in materials with heavy atoms. Hence, we limit our initial set of materials to those having heavy atoms (atomic weight ≥ 65) with bandgap ≤ 0.6 eV and zero magnetic moment value, which results in 4835 materials out of 30000 materials. We calculate the spillage of these materials by comparing wavefunctions from DFT calculations with and without SOC for each material. As introduced by Liu et al.[11], the spin-orbit spillage is a method for comparing wave functions at a given **k**-point with and without SOC, with the goal of identifying band inversion. We calculate the spin-orbit spillage, $\eta(\mathbf{k})$, given by the following equation:

$$\eta(\mathbf{k}) = n_{occ}(\mathbf{k}) - \text{Tr}(P\tilde{P}) \tag{1}$$

where

$P(\mathbf{k}) = \sum_{n=1}^{n_{occ}(\mathbf{k})} |\psi_{n\mathbf{k}}\rangle\langle\psi_{n\mathbf{k}}|$ is the projector onto the occupied wavefunctions without SOC, and $\tilde{P}$ is the same projector with SOC. Here, 'Tr' denotes trace over the occupied bands. We can write the spillage equivalently as:

$$\eta(\mathbf{k}) = n_{occ}(\mathbf{k}) - \sum_{m,n=1}^{n_{occ}(\mathbf{k})} |M_{mn}(\mathbf{k})|^2 \tag{2}$$

where $M_{mn}(\mathbf{k}) = \langle\psi_{m\mathbf{k}}|\tilde{\psi}_{n\mathbf{k}}\rangle$ is the overlap between occupied Bloch functions with and without SOC at the same wave vector **k**. If the SOC does not change the character of the occupied wavefunctions, the spillage will be near zero, while a band inversion will result in a large spillage.

We have extended the original definition of the spillage, which was only defined for insulators, for use in metals by allowing the number of occupied electrons, $n_{occ}(\mathbf{k})$, to vary at different k-points by fixing it to the zero-temperature occupation of the non-SOC calculation. As argued in ref.[11], in the case where the non-SOC calculation gives a trivial insulator and the SOC calculation gives a $Z_2$ topological insulator, the spin-orbit spillage will be at least 2.0 at the k-point(s) where band inversion occurs. At other k-points, or in cases where both materials/calculations are trivial, the spillage is generally much lower. We find that a similar relation holds when SOC drives a crystalline topological phase transition as well. Due to its ease of calculation, without any need for symmetry analysis or dense k-point interpolation, the spillage is an excellent tool for identifying topologically non-trivial insulators and understanding where in k-space the band inversion occurs; although, it is not sufficient to classify the type of TI in general.

In this work, we extend the use of the spillage to cases where one or both of the non-SOC or SOC calculations are metallic, which significantly broadens the class of TMs that we can search for. In the case where either the non-SOC or SOC calculations are metallic, there is no specific spillage value that guarantees a band inversion, but we find empirically that spillages above 0.5 are often due to topologically non-trivial features in metals. For example, in the well-known Dirac material $Na_3Bi$[16], both the SOC and non-SOC calculations have a 3D Dirac cone in the bulk, but the addition of SOC causes the Dirac point to shift off the high symmetry point and onto a symmetry line. Due to this shift, the symmetry of the occupied bands in this region of **k**-space will be different in SOC and non-SOC calculations, which results in a large spillage. Similarly, Weyl points and nodes require SOC, so Weyl materials will always have large spillage near these features. In addition, there are many cases where the addition of SOC can lift a degeneracy, turning a metal into a non-trivial insulator. This gap opening will also generate a large spillage, as again the occupied wavefunctions will change due to SOC. We note that we do not expect the spillage to be useful in identifying TMs where the SOC does not play an important role.

## Results and discussion

To further elucidate how spillage helps in the screening of TMs, we show an example of $Ba_2HgPb$ in Fig. 1. Looking at the band structures shown in Figs. 1a (without SOC) and 1b (with SOC), it is unclear if there as a topological phase transition due to SOC. However, the band structures projected onto the Pb [*p*-orbital] states (Fig.1c and 1d, without and with SOC, respectively), suggest that there is a SOC-induced band-inversion at the L-point. While this band inversion is a strong sign that Ba$_2$HgPb is topologically non-trivial, automating this type of visual analysis based on orbital projections is difficult, as many orbitals can contribute to bands near the Fermi level in complex and low symmetry materials. However, the non-trivial band inversion becomes immediately clear in Fig. 1e, which shows the spillage at each **k**-point, including a peak above 2.0 at the L point. This band inversion at the L point is consistent with Wannier charge center (Fig. S1-S2) calculations that show that Ba$_2$HgPb is a crystalline topological insulator. In the JARVIS-DFT database (https://www.ctcms.nist.gov/~knc6/JVASP.html and https://jarvis.nist.gov ) we have provided plots similar to Fig. 1, including projected band structures, the spillage, and information on band gap, for all materials investigated in this work. In the supplementary materials, we provide examples of these plots for several trivial and non-trivial materials classes (Fig. S3-S7) to elaborate on the spillage criteria. We expect the most promising TMs to have a small number of high peaks in the spillage. We also encounter some metals with high spillage values across the Brillouin zone, which indicates a strong change in the band structure at the Fermi level due to SOC. We find that these materials are less likely to have simple topological features, and instead display many complicated band crossings near the Fermi level. Therefore, after using the maximum spillage as a screening criterion, it is necessary to perform additional analysis to determine the source of the high spillage and classify any topological behavior.

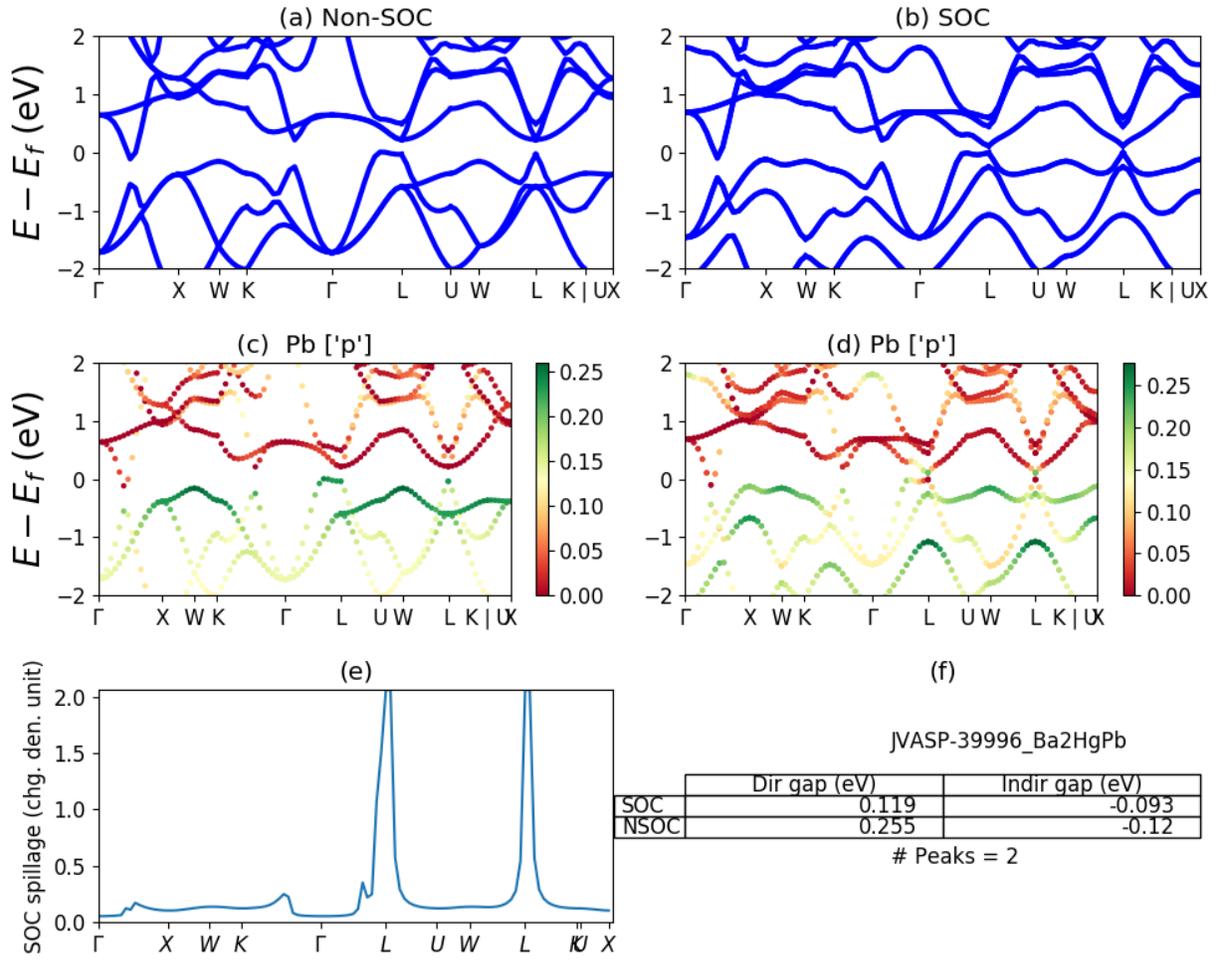

*Fig. 1 a) Non-SOC and b) SOC band structures of Ba$_2$HgPb (JVASP-39996, https://www.ctcms.nist.gov/~knc6/jsmol/JVASP-39996). c) non-SOC and d) SOC projected band structures, projecting onto Pb orbitals. e) Spillage, as a function of k. f) Table of bandgaps and spillage information. Ba$_2$HgPb is an example crystalline topological insulator.*

As discussed earlier, we calculate the spillage values for a large set of low bandgap and high atomic mass materials, with a spillage value of 0.5 as a threshold to screen for potential topologically non-trivial materials. In Fig. 2a, we show the obtained distribution of SOC spillage, with most materials having low spillage even if they had high atomic mass. In Fig 2 b-f) results are shown only for high-spillage materials (spillage ≥ 0.5). Overall the spillage ranges from zero to four.

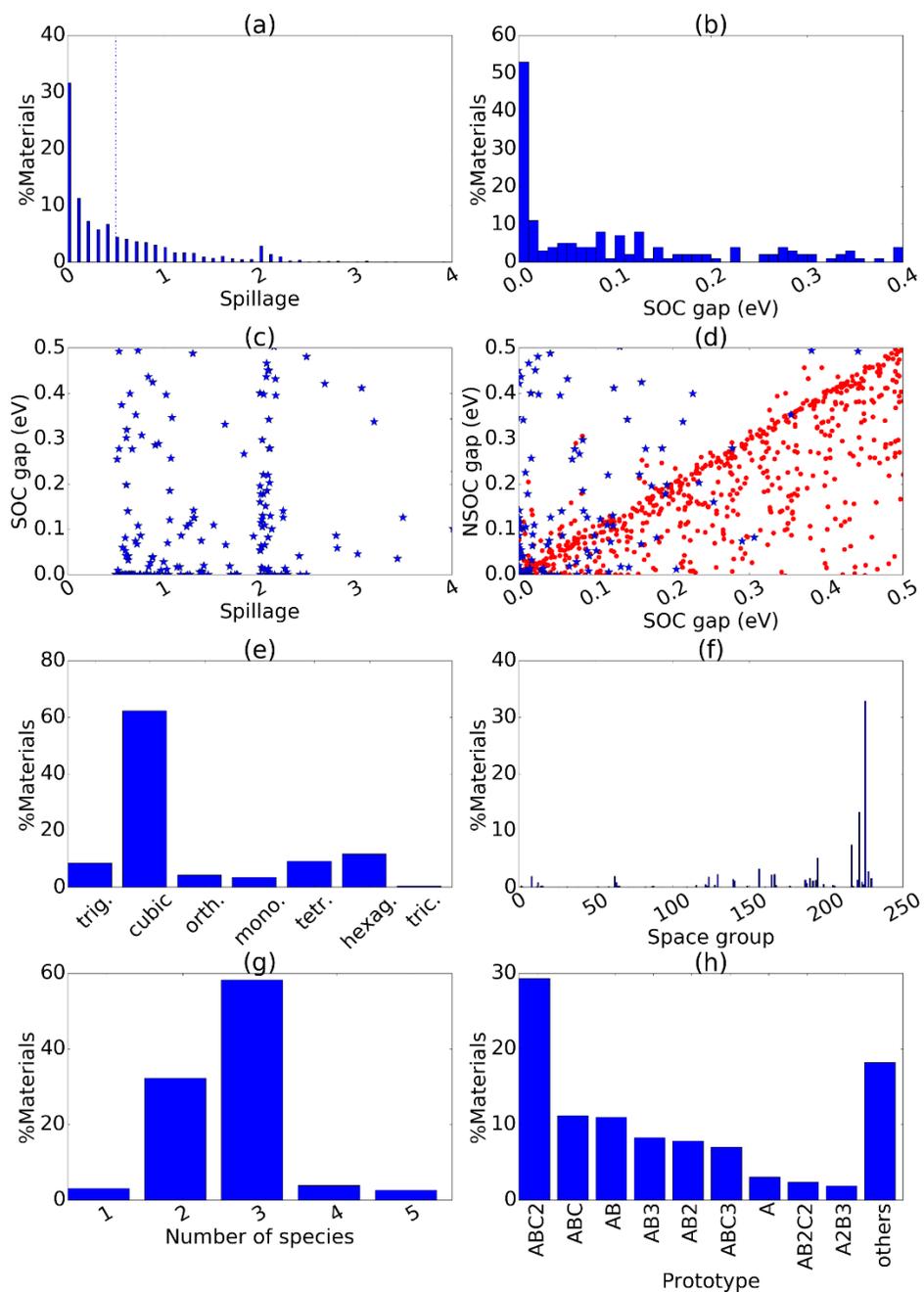

*Fig. 2 Spillage and bandgap related distributions of materials under investigation. Top row: a) Spillage distribution of the all materials in the database, b) SOC bandgap distribution, c) SOC spillage vs SOC bandgaps, d) SOC vs Non-SOC bandgaps. Bottom row: e) crystal system distribution f) space group distribution, g) the number of chemical constituents-based distribution and h) compositional prototype-based distribution. Fig. a show the data for all the materials, while other plots show results for high spillage (≥0.5) materials only.*

In Fig. 2b we analyze the SOC bandgap distribution of all the materials with spillage greater than or equal to 0.5. We find that there are many metallic/semi-metallic materials (about 40% of the high-spillage materials), and wide distribution of band gaps amongst the insulating materials. In Fig. 2c we show the SOC band gap as a function of spillage, and we find clusters of data at spillage values 1 and 2. Our data indicate no correlation between SOC band gap and spillage. It is to be noted that both NSOC and SOC band gaps have the well-known problem of being underestimated as in the case of conventional DFT. However, as this is a systematic effect, it should not affect the comparison between NSOC and SOC values. In Fig. 2d we plot the SOC vs NSOC bandgaps of both high-spillage (spillage ≥ 0.5, blue stars) and low-spillage (red dots) materials. For low spillage materials, there is a fairly strong linear relationship between the NSOC and SOC gaps, indicating that in most cases spin-orbit coupling has little effect on band gaps of topologically trivial materials. However, there are also materials where a large change in band gap occurs, even among low spillage materials. For high spillage materials, there is no relationship between SOC and NSOC band gaps, as expected for materials which have a different band ordering near the Fermi level. This suggests that NSOC band structures should be used with care when studying small gap semiconductors, even if the topological behavior is not specifically desired. In Fig 2e we show the crystal system distribution of all the high spillage materials. We observe that many high spillage materials are cubic, which is in part due to a large number of both half-Heusler alloys and antiperovskites in our DFT database, which are known to display topologically non-trivial behavior[41-43]. There are also many hexagonal materials, including both materials related to the topological insulator $Bi_2Se_3$, as well as hexagonal ABC materials[44-47]. Similar behavior is observed in the space-group distribution in Fig. 2f. Results in Figures 2e and 2f imply that high symmetry in a crystal favors the rise of topologically non-trivial behavior. Moreover, in Fig. 2g we see that most of these materials are ternary; however, there are elemental, binary, and multicomponent non-trivial materials as well. Furthermore, we classify the non-trivial materials in terms of the compositional prototype in Fig. 2h. We also identify the dimensionality of the materials based on the lattice-constant[37] and data-mining approaches[48]. We found 89.61 % materials to be 3D-bulk, 10.02 % to be 2D-bulk, 0.32 % to be 1D-bulk and 0.05% to be 0D materials with high spillage values.

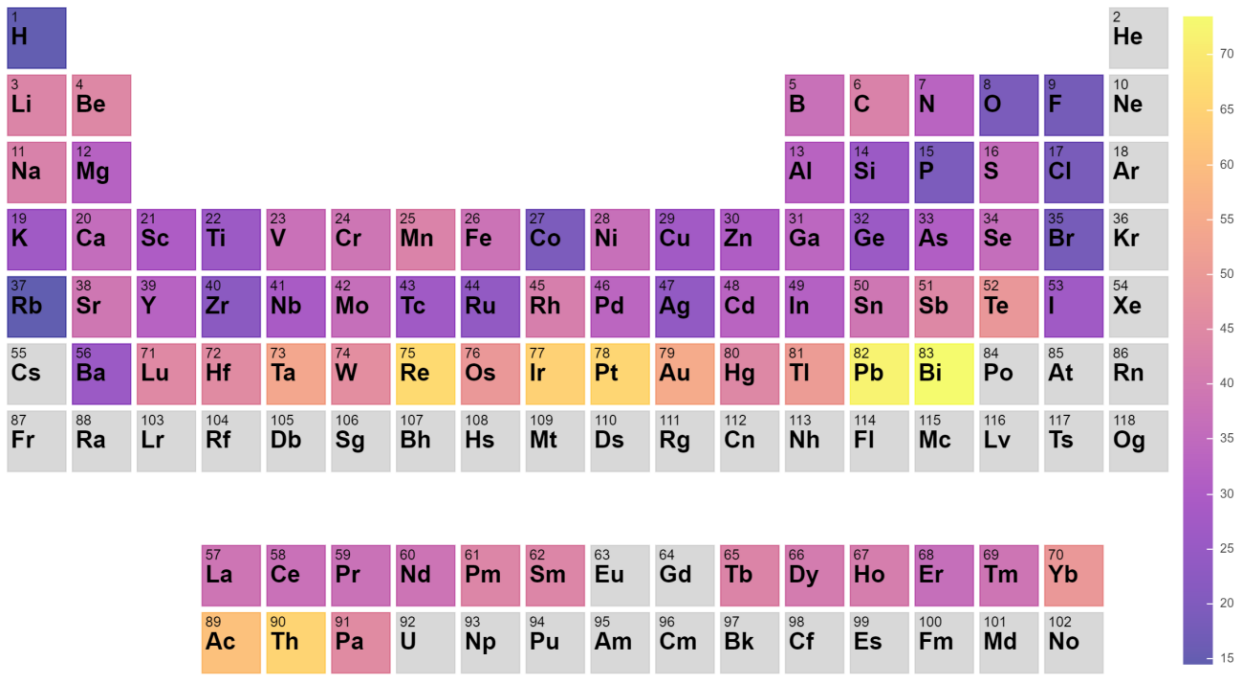

*Fig. 3 Periodic table showing the percentage chance that a compound containing any given element will have spillage above 0.5.*

Next, we analyze the distribution of various high spillage materials as a function of composition. For each element, we calculate the probability that a material in our dataset containing that element has a spillage above 0.5 and is, therefore, a TM candidate. We present the results in the form of a periodic table in Fig. 3. Consistent with known TMs, we observe that materials containing the heavy elements Bi and Pb are by far the likeliest ones to have high spillage. Other high spillage elements are Ta and the 5d elements Re-Pt. To contribute to SOC-induced band inversion, an element must both have significant SOC and contribute to bands located near the Fermi level, which favors heavy elements with moderate electronegativity.

Using the spillage as a screening tool, we greatly reduce the number of candidate TMs to consider, but further analysis is necessary to understand the class of each potential TM. For a few hundred promising candidates, we use Wannier-based tight-binding models to perform efficient Wannier charge center calculations and node finding calculations. We present the results of these calculations in the supplementary information. We characterize 40 strong topological insulators (TI), 7 weak TIs, 24 crystalline topological insulators (CTI), 30 Dirac semimetals, 3 Weyl semimetals, 47 quadratic band-crossing materials. We identify many previously known topological materials of various classes. In addition, we find a variety of promising TMs that, as far as we know, have not been previously reported in the literature. As examples of various types, we present the band structures of a variety of TMs in Fig. 4. As shown in Fig. 4a, we find that PbS in space group P6$_3$/mmc (JVASP-35680) is a strong Z2 topological insulator due to band inversion at Γ. This topology is reflected in the (001) surface band structure (Fig. 4e), which shows a Dirac cone feature at Γ. In Fig. 4b, we show the band structure of LiBiS$_2$ (P4/mmm, JVASP-52332), which is a weak topological insulator due to band inversion at the R point. Fig. 4f shows the surface band structure, including Dirac cone features at M and Z points. In Fig. 4c, we show the band structure of KHgAs (P6$_3$/mmc, JVASP-15801), which is a crystalline topological insulator[49]. The surface band structure (Fig. 4g) shows non-trivial surface bands. In Figs. 4d and 4h, we show the band structure of InSb (JVASP-36123) and GaSb (JVASP-35711) in the P6$_3$mc space group. InSb has a simple bulk Dirac cone feature along the Γ – A line. In contrast, GaSb has a slightly different band ordering that leads to two Dirac cones along Γ – A and complicated Weyl crossings along Γ – K and Γ – M. We find that materials with several crossings near the Fermi level, like GaSb, are relatively common, but simple semimetals like InSb are relatively rare.

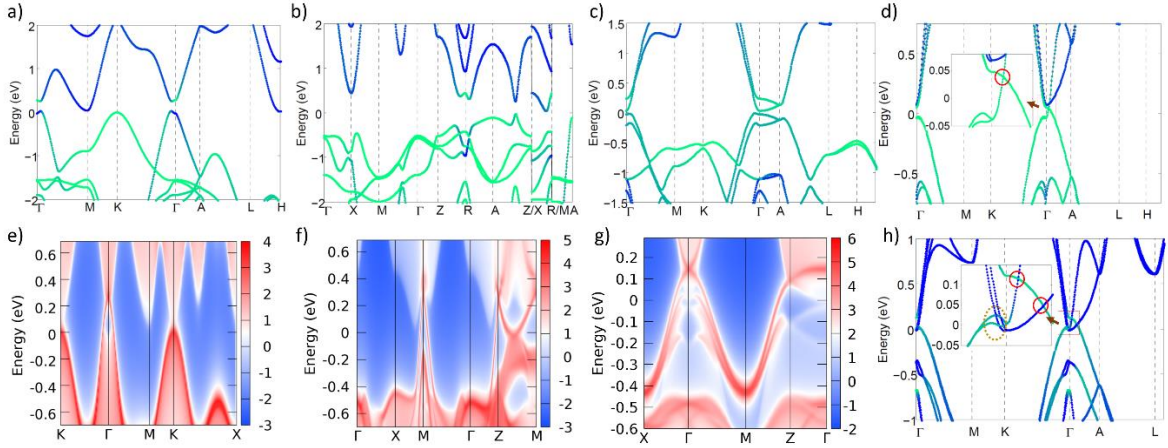

*Fig 4. a) Band structure of PbS, green bands projected onto S orbitals, b) band structure of LiBiS$_2$, projected onto S orbitals, c) band structure of KHgAs, projected onto As orbitals, d) band structure of InSb, projected onto Sb orbitals, e) (001) surface ARPES of PbS f) (100) surface ARPES of LiBiS$_2$, g) (010) slab band of KHgAs, h) band structure of GaSb, projected onto Sb orbitals. Insets in d) and h) show Dirac crossings (red circles) and Weyl crossings (yellow dotted circle).*

## Conclusions

In summary, we have used the spin-orbit spillage criteria to quickly screen a large material database for SOC-related TMs, which we then classified using Wannier charge sheets and band crossing searches. This approach for high-throughput TM discovery has the advantages of being applicable to materials with any symmetry, computationally efficient, and applicable to both strong and weak TI's, as well as many crystalline TIs and semimetals. We found at least 1868 materials to be candidate TMs out 4835 investigated materials, which we present in a database, with the specific topology of hundreds already classified. These new TMs include materials from many crystal structure and chemistries can significantly expand the range of known TMs. Additionally, we show a unique way of extracting information from DFT wavefunctions. We believe that the database can be useful to accelerate the discovery and characterization for TMs and other properties tied to SOC.

## Methods

We perform our spillage-based screening using Vienna Ab-initio Simulation Package (VASP)[50,51] and the PBE functional[52] (with and without SOC). All input structures in this work are relaxed using the OptB88vdW functional[53], which is known to give accurate results for both vdW and not-vdW-bonded materials, as discussed in Ref[38]. However, OptB88vdW+SOC is not publicly available in VASP-package, so we computed spillage using PBE[54]. This approach is reliable because recently Cao *et al.* [55] have shown that band-structures obtained using PBE on vdW-functional-relaxed structure display all the correct features. We also added a comparison of non-spin-orbit coupling bandstructures on same OptB88vdW relaxed geometry using PBE and OptB88vdW functionals for $Bi_2Te_3$ and Ge as examples (Fig. S10).

We calculate the spillage using 1000/atom k-points, as well as by analyzing the spillage along a high-symmetry Brillouin-zone path, with a 600 eV plane-wave cut-off. High values of the spillage are generally restricted to regions of k-space where band inversion occurs, so as long as the k-point grid is dense enough to locate these regions, the spillage can be calculated with a sparse k-point grid (see Table S1). For topological insulators, band inversion generally happens at high-symmetry points, which are included in any typical k-point grid. For semimetals, where topological features can occur at arbitrary k-points, it is necessary to search on a grid; however, we allow tolerance for what we consider high spillage (spillage ≥ 0.5) in part to avoid the need to hit the highest spillage point exactly in order to identify a potentially interesting material.

Analysis of the spillage can in some cases help identify the type of topological phase. For example, strong topological insulators with $Bi_2Se_3$-like structures have band inversion at the Γ-point only, and the spillage reflects this pattern, with spillage above 2 at Γ-point, but dropping to near zero when moving along away from Γ. In contrast, for the weak topological insulator BiSe (JVASP-5410), which has a strongly two-dimensional band structure, the bands are inverted along the entire Γ-A line, which is also reflected in the spillage. However, the spillage cannot be used to identify the topological phase in all cases. For example, $Tl_8Sn_2Te_6$ (JVASP-9377) does have a SOC-induced band inversion at Γ, however, more detailed analysis (as discussed below) reveals that the resulting band structure is topologically trivial. False positive examples like this one are very rare. These examples are presented in the supplementary materials (Fig. S3, S8 and S9).

Therefore, in order to investigate 289 candidate materials further, we used Quantum Espresso (QE)[56] with norm-conserving pseudopotentials[57,58] and Wannier90[59,60] to generate first principles tight-binding models. Using WannierTools[12], we then performed Wannier change center (Wilson loop)[61,62] calculations of topological invariants, a numerical search for band crossings, and surface state calculations.

## Data availability

The electronic structure data is available at the JARVIS-DFT website: https://www.ctcms.nist.gov/~knc6/JVASP.html and http://jarvis.nist.gov . We also provide a JSON file with the spillage, bandgap (with and without spin-orbit coupling) and topological material type information at the figshare link: **https://doi.org/10.6084/m9.figshare.7594571.v1** .

## Contributions

KC and KG jointly developed the method. KC and KG carried out the high-throughput DFT calculations using VASP and QE packages respectively. FT helped in writing the manuscript.

## Competing interests

The authors declare no competing interests.